\documentstyle[12pt]{article}
\begin{document}
\begin{center}
{\large \bf 
Multifractal Measures for the Yen-Dollar Exchange Rate 

\vspace*{.5in}

\normalsize 
Kyungsik Kim$^{*}$, Seong-Min Yoon$^{\dagger}$ and Jum-Soo Choi  \\

\vspace*{.1in}

{\em 
Department of Physics, Pukyong National University,\\
Pusan 608-737, Korea\\
$^{\dagger}$Division of Economics, Pukyong National University,\\
 Pusan 608-737, Korea}\\ 

}
\end{center}


%

\hfill\\
%
%
%
\baselineskip 24pt
\hfill\\   
We study the tick dynamical behavior of the yen-dollar exchange rate using the rescaled 
range analysis in financial market. It is found that the multifractal Hurst exponents with 
the short and long-run memory effects can be obtained from the yen-dollar exchange rate.
This exists one crossover for the Hurst exponents 
at charateristic time scales, while the bond futures exists no crossover. 
Particularly, it is shown that the probability distribution of the yen-dollar exchange rate has 
one form of the Lorentz distribution rather than fat-tailed properties, which is similar to that of
for the won-dollar exchange rate.

%
\hfill
\vskip 2mm 
\hfill\\
$^{*}$ Email: kskim@pknu.ac.kr; Tel.: +82-51-620-6354; Fax: +82-51-611-6357  \\

\newpage


Recently, the outstanding topics in econophysics have mainly included 
the price changes in open market $[1, 2]$, the distribution of income of companies,
the scaling relation of size fluctuations of companies, the financial analysis of foreign 
exchange rates $[3]$, the tick data analysis of bond futures $[4]$, the herd behavior of financial markets $[5]$,
the self-organized segregation, and the minority-majority game $[6]$.  
In particular, the essential problems with fluctuations have particularly led to a better understanding for
the scaling properties based on methods and approaches in scientific fields.
It was discussed in the previous work $[3]$ 
that the price fluctuations follow the anomalous power law
from the stochastic time evolution equation, which is represented in terms of the Langevin-type equation. 
Furthermore, the power law distribution, the stretched exponential
distribution, and the fat-tailed distribution have showed the functional properties
from the numerical results obtained in diverse econophysical systems.

To measure the multifractals of dynamical dissipative systems, the generalized dimension 
and the spectrum have effectively used to calculate the trajectory of chaotic attractors 
that may be classified by the type and number of the unstable periodic orbits.
Many attempts $[7-9]$ to compute these statistical quantities have primarily presented from the box-counting method.   
We have used the box-counting method to analyze precisely generalized dimensions and scaling exponents 
for the mountain height and the sea-bottom depth $[10]$.
For the standard analysis, since there exists notably no statistical correlations between observations,     
the R/S analysis$[11]$ has extended to distinguish the random time series from correlated ones.

Scalas $et$ $al.$ $[4]$ have studied the correlation function for bond walks from the time series of 
Buoni del tesoro Poliennali futures exchanged at 
the London International Financial Futures and options Exchange(LIFFE).  
They have discussed that the continuous-time random walk theory $[12]$ is sucessfully applied to the dynamical behavior
of empirical scaling laws by a set of tick-by-tick data in financial markets.
Mainardi $et$ $al.$ $[13]$ have also argued on the waiting-time distribution for bond futures traded at 
LIFFE.
The theoretical and numerical arguments for the volume of bond futures 
were presented at Korean Futures Exchange market $[14]$.
The studies of multifractals in financial markets have explored up to now,
but it is of fundamental importance 
to treat with the multifractal nature of prices for the yen-dollar exchange rate.
In this paper, we treat with the generic multifractal behavior for tick data of prices 
using the R/S analysis for the yen-dollar exchange rate. 
The result obtained is also compared with that of the won-dollar exchange rate, and
the multifractal Hurst exponents, the price-price correlation 
function, and the probability distribution of returns are particularly discussed
with long-run memory effects.
Our result obtained is also compared with that of the won-dollar exchange rate. 

To quantify the Hurst exponents, we introduce the R/S analysis method that is generally contributed 
to estimate the multifractals $[15,16]$.
At first we suppose a price time series of length $n$ given by
$ \lbrace p(t_1 ),  p(t_2 ), ...,p (t_n )  \rbrace$,
and the price $\tau$-returns $r(\tau)$ having time scale $\tau$ and length $n$ that is represented in terms of
$r(\tau) = \lbrace r_1 (\tau),  r_2 (\tau), ...,r_n (\tau) \rbrace$,
with $r_i (\tau)=\ln p(t_i +\tau) - \ln p(t_i )$.
For simplicity, after dividing the time series or returns into $N$ subseries of length $M$, we label each subseries 
$E_{M,d} (\tau )=\lbrace r_{1,d} (\tau ),r_{2,d} (\tau ),..., r_{M,d} (\tau ) \rbrace$, with $d=1,2,...,N $. 
Then, the deviation $ D_{M,d} (\tau )$, i.e., the differences between $ r_{M,d} (\tau )$ and $ {\bar{r}}_{M,d} (\tau )$
for all $M$, can be defined from the mean of returns ${\bar{r}}_{M,d} (\tau )$
as
\begin{equation}
D_{M,d} (\tau )= \sum_{k=1}^{M}( r_{k,d} (\tau )- {\bar{r}}_{M,d} (\tau )).
\label{eq:f6}
\end{equation} 
The hierachical average value $(R/S)_M (\tau )$ represented the rescaled and normalized relation between 
the subseries $R_{M,d} (\tau )$ and the standard deviation $S_{M,d} (\tau )$ becomes  
\begin{equation}
(R/S)_M (\tau )= \frac{1}{N} \sum_{d=1}^{N} \frac{R_{M,d} (\tau )}{S_{M,d} (\tau )}
 \propto M^{H(\tau )},
\label{eq:h8}
\end{equation} 
where $H(\tau )$ is called the Hurst exponent, and
the statistical quantities $ R_{M,d} (\tau )$ and $S_{M,d} (\tau )$ are, respectively, given by
%
\begin{eqnarray}
R_{M,d} (\tau ) &=& max \lbrace D_{1,d} (\tau ), D_{2,d} (\tau ),...,D_{M,d} (\tau ) \rbrace \nonumber \\
&&-min  \lbrace D_{1,d} (\tau ), D_{2,d} (\tau ),...,D_{M,d} (\tau ) \rbrace
\label{eq:g7}
\end{eqnarray}
and
\begin{equation}
S_{M,d} (\tau )=
[ \frac{1}{M} \sum_{k=1}^{M}( r_{k,d} (\tau )- {\bar{r}}_{M,d} (\tau ))^2]^{\frac{1}{2}}.
\label{eq:e5}
\end{equation} 
%
%

For more than one decade,
several methods have suggested in order to investigate the multifractal properties systematically. 
From the multifractality of self-affine fractals $[7-9]$, the $q$-th price-price correlation function $F_q (\tau)$ 
takes the form
\begin{equation}
F_q (\tau)=<|p(t+\tau)-p(t)|^q >  \propto {\tau}^{qH_q }, 
\label{eq:j10}
\end{equation} 
where $\tau$ is the time lag, $H_q$ is the generalized $q$th-order Hurst exponent, and 
the angular brackets denote a statistical average over time.
When our simulation is performed on the price $p(t)$,
a nontrivial multi-affine spectrum can be obtained as $H_q$ varies with $q$.
This has exploited in the multifractal method and the large fluctuation
effects in the dynamical behavior of the price can be explored from Eq.$(5)$.
In our scheme, we will make use of Eqs.$(2)$ and $(5)$ to compute the multifractal features of
prices, and the mathematical techniques discussed lead us to more general results.

For characteristic analysis of the yen-dollar exchange rate,
we will present in detail numerical data of Hurst exponents from the results of R/S analysis.
Although we extend to find other statistical quantities via computer simulation studies of returns, 
we resrict ourselves to estimate the generalized $q$th-order Hurst exponents in the price-price correlation 
function and the form of the probability distribution of returns.
In this paper, we introduce the price time series for the yen-dollar exchange rate, in which 
the time step between ticks is evoluted for one day and the tick data for the yen-dollar exchange rate were 
taken from January $1971$ to June $2003$.  
The Hurst exponents are obtained numerically from the results of R/S analysis given by Eq.$(2)$, as summarized in Table $1$.
Fig.$1$ shows the tick data of returns $R(t)$ as a function $\tau=1$(one day) for the yen-dollar exchange rate, 
and the Hurst exponents for the yen-dollar exchange rate are $H(\tau= 1)$=$0.6513$, 
as plotted in Fig.$2$. 
It is found that our Hurst values are significantly different from 
the random walk with $H=0.5$, which 
this process are located in the persistence region similar to those of the crude oil prices $[16]$.
Especially, it may be expected that the Hurst exponent is taken anomalously to be 
near $1$ as the time series proceeds with long-run memory effects.
The crossover in the Hurst exponent $ H(\tau)$ is not existed,
while $H(\tau )$ from our tick data is similarly found to have the existence of crossovers 
at characteristc time $\tau=9$($\tau=7$ and $35$) for the won-dollar exchange rate(the KOSPI) $[17]$.

For the sake of concreteness, we perform the numerical study of Eq.$(5)$ in order to analyze the 
generalized $q$th-order Hurst exponents in the price-price correlation function $F_q (\tau)$.
Table $1$ includes the values of the generalized $q$th-order Hurst exponent $H_q$ in the price-price correlation function 
for the yen-dollar exchange rate. 
Especially, the generalized Hurst exponent is taken to be near $0.6216$ as $q\to 1$,
and the values log$(F_q /q)$ for $q=1,2,...,6$ are plotted in
Fig.$3$ for the the yen-dollar exchange rate.
The probability distribution of returns is well consistent with a Lorentz distribution
different from fat-tailed properties, as shown in Figs.$4$ and $5$.

In summary, we have presented the multifractal measures from the dynamical behavior 
of prices using the R/S analysis for the yen-dollar exchange rate.
The multifractal Hurst exponents, the generalized $q$th-order Hurst exponent, and 
the form of the probability distribution have discussed with long-run memory effects.
Since Hurst exponents are larger than $0.5$ through R/S analysis, our case for time series of prices is
the persistent process.
It is apparent from our data of the Hurst exponent $H(\tau )$ that
the existence of crossovers is similar to that of other result $[16]$.
Moreover, it is found that the probability distribution for all returns is well consistent with a Lorentz distribution.
Since it supports to carry out the dynamical behavior in our stock and foreign exchange markets,
our analysis would assure that it is able to capture the essential multifractal properties in our present
result. 
In future, our result will be applied to extensively investigate the other tick data 
in Korean financial markets and compared with other calculations transacted in other nations in detail.

\vskip 5mm
\noindent
{\bf ACKNOWLEDGMENT}
\hfill\\
This work was supported by Grant No.R01-2000-000-00061-0 from the Basic Research 
Program of the Korea Science and Engineering Foundation.
%
%

%

%
%
\newpage
\vskip 10mm
\begin{center}
{\bf FIGURE  CAPTIONS}
\end{center}

\vspace {5mm}



\noindent
Fig. $1$.  Plot of the tick data for the yen-dollar exchange rate, where one time step is the transaction
time evoluted for one day.
This continuous tick data were taken from January $1971$ to June $2003$. 
\vspace {10mm}

\noindent
Fig. $2$.  Log-log plot of $R/S(\tau)$ at $\tau=1$ for the yen-dollar exchange rate.
\vspace {10mm}

\noindent
Fig. $3$.  Plot of The $q$-th price-price correlation function $F_q (\tau)$ of the time interval $\tau$ for 
the KOSPI, where the value of slopes is summarized in Table $1$.
\vspace {10mm}

\noindent
Fig. $4$.  The probability distribution of returns for the KOSPI. 
the dot line is represented in terms of a Lorentz distribution, 
i.e. $ P(r) =$$  \frac{2b}{\pi} \frac{a}{{r}^2 + a^2} $, 
where $ a=7.0\times 10^{-3} $ and $ b = 9.0 \times 10^{-3}$
for the yen-dollar exchange rate.

\vspace {10mm}
\noindent
Fig. $5$.  The probability distribution of all returns for the yen-dollar exchange rate. The dashed and solid lines show
the Gaussian and Lorentz distributions, respectively, where the right(left) tail denotes the postive
(negative) returns region. 
\vspace {10mm}
\newpage
\vskip 10mm

\begin{center}
{\bf TABLE  CAPTIONS}
\end{center}
\vspace {5mm}
\noindent
Table $1$.  Summary of values of the Hurst exponent $H(\tau)$ and
the generalized $q$th-order Hurst exponent $H_q$ for the yen-dollar exchange rate(YDER),
th won-dollar exchange rate(WDER).
\vspace {3mm}\\
\begin{tabular}{l|c|rr}\hline\hline
          &  $H(\tau)$               &  $H_q $ \\  \hline                           
YDER      &  $H(\tau= 1) = 0.6513$   &  $H_1 =$$0.6216   $  &  $H_4 =$$0.4148   $      \\
          &  $H(\tau=15) = 0.5710$   &  $H_2 =$$0.5688   $  &  $H_5 =$$0.3486   $      \\
          &  $H(\tau=25) = 0.5885$   &  $H_3 =$$0.4956   $  &  $H_6 =$$0.2996   $    \\  \hline
%
WDER      &  $H(\tau= 1) = 0.6886$ &  $H_1 =$$0.6535$   &  $H_4 =$$0.4307 $      \\
          &  $H(\tau= 5) = 0.7283$ &  $H_2 =$$0.5614 $  &  $H_5 =$$0.3914 $      \\
          &  $H(\tau=25) = 0.7372$ &  $H_3 =$$0.4859 $  &  $H_6 =$$0.3629$      \\ \hline \hline
%
\end{tabular}
\vspace {5mm}
%
%

%

\end{document}